\documentclass[a4paper, 12pt]{article}

\usepackage{fullpage}
\usepackage[utf8]{inputenc}
\usepackage[T1]{fontenc}
\usepackage{tikz}
\usetikzlibrary{quantikz}
\usepackage[title]{appendix}

\usepackage{graphics}
\usepackage{cite}
\usepackage{braket}
\usepackage{authblk}
\usepackage{titlesec}
\PassOptionsToPackage{hyphens}{url}\usepackage{hyperref}
\usepackage{hyperref}

\title{Quantum amplitude estimation with error mitigation for time-evolving probabilistic networks}

\author[1]{M.C. Braun}
\author[1]{T. Decker\footnote{For contact email: thomas.decker@jos-quantum.de or sven.kerstan@jos-quantum.de. Authors are listed in alphabetical order.}}
\author[1]{N. Hegemann}
\author[1]{S.F. Kerstan}
\author[2]{C. Maier}
\author[2]{J. Ulmanis}

\affil[1]{JoS QUANTUM GmbH, c/o Tech Quartier, Platz der Einheit 2, \protect\\60327 Frankfurt am Main, Germany}
\affil[2]{Alpine Quantum Technologies GmbH, Technikerstrasse 17 / 1, \protect\\A-6020 Innsbruck, Austria}

\begin{document}

\maketitle

\vspace{-0.5cm}
\begin{abstract}
We present a method to model a discretized time evolution of probabilistic networks on gate-based quantum computers. We consider networks of nodes, where each node can be in one of two states: good or failed. In each time step, probabilities are assigned for each node to fail (switch from good to failed) or to recover (switch from failed to good). Furthermore, probabilities are assigned for failing nodes to trigger the failure of other, good nodes. 
Our method can evaluate arbitrary network topologies for any number of time steps. We can therefore model events such as cascaded failure and avalanche effects which are inherent to financial networks, payment and supply chain networks, power grids, telecommunication networks and others.
Using quantum amplitude estimation techniques, we are able to estimate the probability of any configuration for any set of nodes over time. This allows us, for example, to determine the probability of the first node to be in the good state after the last time step, without the necessity to track intermediate states.
We present the results of a low-depth quantum amplitude estimation on a simulator with a realistic noise model. We also present the results for running this example on the AQT quantum computer system PINE.
Finally, we introduce an error model that allows us to improve the results from the simulator and from the experiments on the PINE system.
\end{abstract}

\section{Introduction}
The analysis of networks and their time evolution has applications in several industries, for example in financial risk management where counterparty risk needs to be evaluated to calculate regulatory capital requirements. 
Value at risk (VaR) or expected shortfall (ES) are common risk measures that need to be calculated by banks, asset managers, regulators and insurers on a regular basis. 
For example, we can think of a network of companies and banks that have financial obligations like credit contracts or that have the same assets on the balance sheet. A bank needs to control its overall credit exposure by estimating the likelihood of each counterparty defaulting, called probability of default. As the financial network is highly correlated, systemic risk needs to be considered as cascading effects can destabilize the system or parts of it~\cite{Nature2012, complexNetworks2003, complexFN2016}.

In this publication, we present a method to model a discretized time evolution of probabilistic networks on gate-based quantum computers. First, we present the model that evolves over time in section~\ref{section model} and then extend it to a quantum version in section~\ref{section quantum}. The quantum version can be run on gate-based quantum computers, which might lead to computational benefits, when the quantum computing hardware is available in the required size and precision. 
Compared to our previous work for the sensitivity analysis of business risk models~\cite{JoS2021}, the model for networks and the model for business risks have in common that there are intrinsic and trigger probabilities. The extended network model, which we present in this paper, allows for arbitrary network topologies including cycles as the states in each time step depend only on the states of the previous time steps. Furthermore, each node comes with a recovery probability that allows a node to recover after a failure with a certain probability. To evaluate probabilities of default, Monte Carlo methods can be used to evaluate the model by randomly sampling results. The error of classical Monte Carlo methods decreases with $\sqrt{1/N}$ for $N$ samples. Thus calculating results with a $10$ times higher accuracy, requires $100$ times the computational effort. Depending on the size of the network this can get computational prohibitive.

We can use Monte Carlo methods on quantum computers, which refer to a quantum algorithm called quantum amplitude estimation (QAE)~\cite{QAE}, to calculate risk measures for the probability of the failure of one node or a group of nodes. Various applications of QAE have been shown within finance, e.g. risk analysis~\cite{RiskIBM, JoS2021}, the pricing of financial derivatives~\cite{OptionPricing2018, OptionPricing2020} and many more~\cite{Overview2020, GoldOverview2020}. Using QAE would achieve a quadratic speedup compared to its classical counterpart and therefore requires only $10$ times the computational effort for $10$ times higher accuracy.

The standard amplitude estimation procedure~\cite{QAE} has hardware requirements that are challenging for current quantum devices due to noise. An example for a real world business problem was analyzed in ~\cite{JoS2021}. For that example, it was estimated that the successful execution on a quantum computer would require it to perform at least $100$ million gate operations before a single error occurs, on average.
Furthermore, several assessments of the necessary overhead for quantum error correction have led to discouraging predictions for what would be required to achieve solutions that are faster or more cost-efficient than classical algorithms~\cite{Google2021}. Reducing the requirements of a QAE is currently an active area of research~\cite{QAEwoPhase,MLQAE,QAC,iQAE,LDQAE,QMCI,VQAE} offering the possibility to implement QAE with lower requirements towards hardware fidelity~\cite{lowQAE}.

We assess the viability of the analysis of a small network model with a variant of low-depth QAE methods in section~\ref{section QAE}. We introduce a noise model, which can be fitted to measurement results by gradient descent methods. For example, this allows us to improve the results of QAE for determining the probabilities of the failure of nodes in a network. We implement amplitude estimation for the quantum model on a simulator with a realistic hardware noise model as well as on real hardware from AQT~\footnote{AQT's quantum computer system PINE.}. The result shows, that low-depth versions of QAE are currently feasible for small networks with one or two nodes and up to three or four time steps.

To summarize, the main contributions of this paper are the following:
\begin{enumerate}
\item{The implementation of the time evolution of a probabilistic network model as a quantum circuit.}
\item{The execution of low-depth quantum amplitude estimation (QAE) of the model on a physical quantum computer: AQT's PINE system.}
\item{Evaluating the low depth QAE results with a simple noise model to improve the estimation of the probabilities, which we want to determine.}
\end{enumerate}

\section{The network model}\label{section model}
In this section, we define the network model. For a simple example, we show the results of a direct calculation of probabilities and compare these to the results of classical Monte Carlo evaluations. The exact and Monte Carlo evaluations for the network model are implemented in the Pygrnd library~\cite{pygrnd} and both methods are explained in detail in appendix~\ref{appendix 1} and~\ref{appendix 2}.

\subsection{Definition of the model}\label{section definition model}
We model a network by a set $N=\{1, \ldots, k\}$ of $k$ nodes. We consider $T$ time steps and for each time step $t \in \{1, \ldots, T\}$ a node $n\in N$ can have the state $0$ or $1$.
We use the functions $s_n(t)$ for $n \in N$ and $t\in\{1,\ldots,T\}$ to describe the state of node $n$ in time step $t$. We set $s_n(t)=0$ if node $n \in N$ is good and $s_n(t)=1$ if the node has failed. A configuration $c \in \{0,1\}^k$ at a time step is the sequence
$c=(s_k(t),\ldots, s_1(t))$ of the states of all nodes~\footnote{We use this order of elements to be consistent with the qubit ordering of Qiskit.}. We write $p_c(t)$ for the probability for configuration $c$ in time step $t$.

For each node $n$, we have the intrinsic probability $p^{\rm fail}_n$ to fail in a time step if the node was good in the previous time step. Each node has also the probability $p^{\rm recover}_n$ to recover if the state was $1$ in the previous time step. Furthermore, if a node $m \in N$ has failed in time step $t$ then it might trigger the default of another node $n \in N$ in the next time step with a certain probability. We write this probability as $p^{\rm trigger}_{m,n}$.

We now complete the description of the time evolution of the system. When we assume that a node $n \in N$ has failed in time step $t$, then we only consider the recovery probability for setting $s_n(t+1)$ and no intrinsic or trigger probabilities are taken into account.
If a node $n\in N$ is good in time step $t$, then we set it to the failed state with the probability $p^{\rm fail}_n$. We also iterate through all nodes $m \in N$ in the model with $p^{\rm trigger}_{m,n}>0$ and we set $n$ to the failed state with this probability. If in this procedure the state turns to the failed state, then we do not set it to good again in this time step.

Note that the states in a time step depend only on the states of the previous time step. This separation of time steps allows us to have arbitrary dependencies between the nodes without ambiguities, e.g. it is possible that two nodes can trigger each other. We assume that time step $t=1$ is the initial time step and that all nodes have the state $1$ with the corresponding intrinsic probability of failure. This can be seen as the result of a time step when for $t=0$ all nodes are in state $0$.

\subsection{Network example}\label{subsection example}
In this section, we define the example that we use for most of the following discussions. We consider a model with two nodes and three time steps, i.e. we have $k=2$ and $T=3$. The probabilities of the model are defined in table~\ref{tables parameters example}. We would like to find the probabilities for $p_c(t)$ for the configurations $c\in \{00,10,01,11\}$ and time step $t=3$. These are calculated in the following sections with a classical calculation of the probabilities and Monte Carlo methods and with a quantum circuit.
\begin{table}[h]
\centering
\caption{The parameters of the example with 2 nodes.}
\vspace{0.5cm}
\begin{tabular}{c|cc}
node&$p_n^{\rm fail}$ & $p_n^{\rm recover}$  \\ \hline
1&0.2&0.3\\
2&0.7&0.8
\end{tabular}
\quad
\begin{tabular}{c|cc}
$p^{\rm trigger}_{m,n}$ & $n=1$ & $n=2$  \\ \hline
$m=1$&-&$0.2$\\
$m=2$&$0.8$&-
\end{tabular}
\label{tables parameters example}
\end{table}

\subsection{Classical evaluation of the model}\label{subsection classical model}
Following the rules of section~\ref{section definition model}, we can evaluate the probabilities of configurations in a time step with an exact calculation. The method is described in appendix~\ref{appendix 1}. Although the implemented method is not efficient and cannot be used for models with many nodes and time steps, we can use it to determine the probabilities of configurations for small examples. This allows us to assess the precision of the results of Monte Carlo methods in section~\ref{subsection monte carlo} and to evaluate the performance of the quantum circuits that we define in section~\ref{section quantum}.
The results\footnote{In this paper, we round all numerical values to three decimal places to simplify the notation.} of the evaluation for the example of section~\ref{subsection example} with 2 nodes for all time steps up to $3$ are shown in table~\ref{table 2 nodes}.
\begin{table}[h]
\centering
\caption{Time evolution of the probabilities of the configurations of 
the network model from section~\ref{subsection example} with 2 nodes.}
\vspace{0.5cm}
\begin{tabular}{c|cccc}
t&$p_{00}(t)$&$p_{10}(t)$&$p_{01}(t)$&$p_{11}(t)$\\ \hline
0&1.0&0.0&0.0&0.0\\
1&0.240&0.560&0.060&0.140\\
2&0.167&0.174&0.479&0.179\\
3&0.140&0.219&0.308&0.333
\end{tabular}
\label{table 2 nodes}
\end{table}

\subsection{Monte Carlo evaluation of the model}\label{subsection monte carlo}
Besides the exact evaluation method of section~\ref{subsection classical model}, we can obtain the probabilities $p_c(t)$ for each possible configuration $c\in \{0,1\}^k$ after $t$ time steps by Monte Carlo simulations of the model. A possible implementation is outlined in appendix~\ref{appendix 2}.

\begin{figure}[h]
\centering
\includegraphics[width=1.0\textwidth]{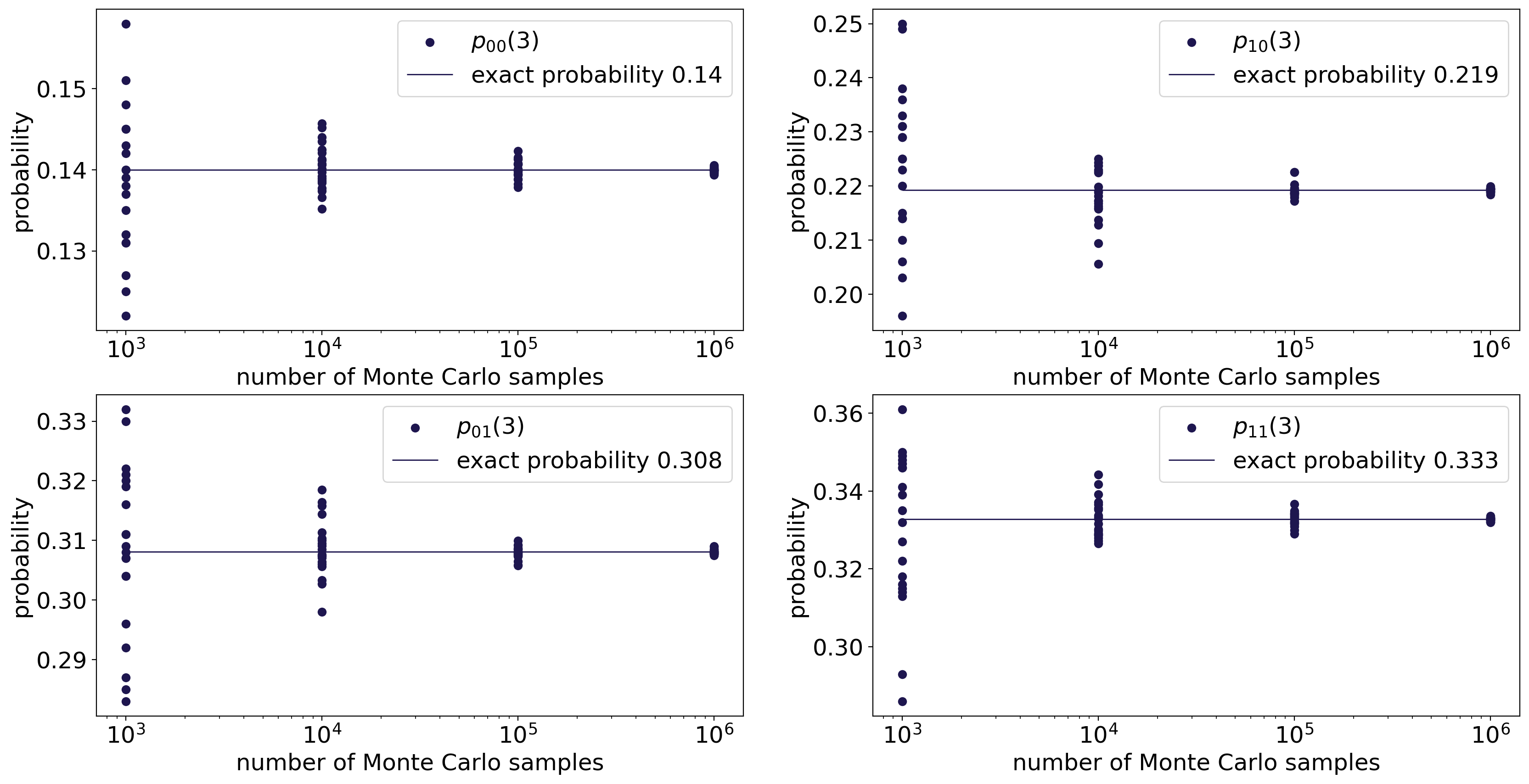}
\caption{Results of Monte Carlo simulations for the example of section~\ref{subsection example}. The probabilities of the configurations calculated by the classical method can be found in table~\ref{table 2 nodes}.}
\label{fig:MC example}
\end{figure}

The results of several Monte Carlo evaluations for the example of section~\ref{subsection example} can be seen in figure~\ref{fig:MC example}. It shows the results for calculating $p_c(t)$ for all configurations $c \in \{0,1\}^2$ for time step $t=3$. The Monte Carlo evaluations were done with $10^3$, $10^4$, $10^5$ and $10^6$ runs. For each number of runs, we performed 20 independent calculations and the stability of the results depending on the number of runs can be seen in the spread of the data points. The horizontal lines correspond to the probabilities that are calculated by the method described in section~\ref{subsection classical model}.

\section{Time evolution of networks as quantum circuits}\label{section quantum}
The network model of section~\ref{section model} can be used to construct quantum circuits, which reproduce the probabilities $p_c(t)$ for all configurations $c\in\{0,1\}^k$ and for all time steps $t$. The definition of the networks and the transitions between time steps allows us to build quantum circuits for any network topology, including cycles in the dependency graph of the nodes. 

\subsection{Construction of quantum circuit for a model}
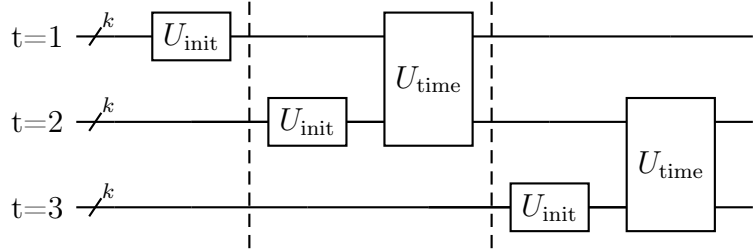
\begin{figure}[h]
\centering
\begin{quantikz}
\lstick{t=1}&\qwbundle{k}& \gate{U_{\rm init}}\slice[style=black]{}& \qw                   & \gate[2]{U_{\rm time}}\slice[style=black]{}& \qw                    & \qw            &\qw\\
\lstick{t=2}&\qwbundle{k}& \qw                        & \gate{U_{\rm init}}\qw& \qw                           & \qw                    & \gate[2]{U_{\rm time}} &\qw\\
\lstick{t=3}&\qwbundle{k}& \qw                        & \qw                   & \qw                           & \gate{U_{\rm init}}\qw & \qw                    &\qw 
\end{quantikz}
\caption{A circuit that implements three time steps after the initial time step for a network with $k$ nodes, i.e. we have $t=3$. The unitary $U_{\rm init}$ initializes the $k$ qubits corresponding to the intrinsic probabilities $p_n^{\rm fail}$ to fail for each node. The unitary $U_{\rm time}$ implements the operator for one time step and it generates the states for the output on the lower register.}
\label{circuit 3 timesteps}
\end{figure}

A network with $k$ nodes is represented by $k$ qubits for each time step. We write $q_n(t)$ for the qubit corresponding to node $n \in N$ in time step $t$. The states $|0\rangle$ and $|1\rangle$ of the qubit correspond directly to the states $0$ and $1$ of a node. We consider
an iterative construction with an operator $U_{\rm init}$ to initialize a register and an operator $U_{\rm time}$ that connects two registers for consecutive time steps $t$ and $t+1$ and that calculates the probability distribution for $t+1$.
The structure of the circuit corresponding to a model with $k$ nodes and $3$ time steps after the initial time step is shown in figure~\ref{circuit 3 timesteps}.

\begin{figure}[h]
\centering
\begin{quantikz}[row sep=1.7cm]
\lstick{$q_1(t)$} & \gate[2,style={inner ysep=2.5pt}]{U_{\rm init}} &\qw \\
\lstick{$q_k(t)$} &  &\qw
\end{quantikz}
=
\begin{quantikz}
\lstick{$q_1(t)$} & \gate{R_y(2\,{\rm arcsin\sqrt{p_1^{\rm fail}}})} &\qw \\
    & \vdots                                         &    \\
\lstick{$q_k(t)$} & \gate{R_y(2\,{\rm arcsin\sqrt{p_k^{\rm fail}}})} &\qw
\end{quantikz}
\caption{The implementation of $U_{\rm init}$ on the register with $k$ qubits for time step $t$. When we measure the generated state of this gate then we obtain the result $1$ with  probability $p_k^{\rm fail}$ to fail for each node $k$. We use the notation of Qiskit for the gates $R_y(\theta)$.}
\label{circuit init}
\end{figure}
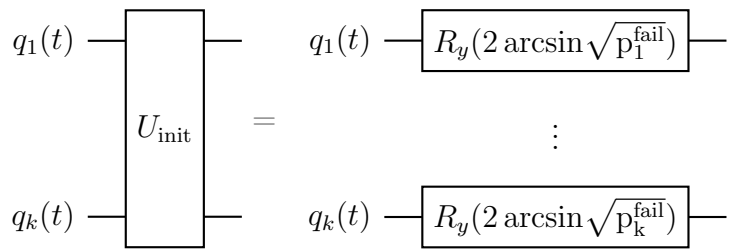

The operator $U_{\rm init}$ initializes a register for a time step with the intrinsic probabilities $p_n^{\rm fail}$ for each node $n \in N$. After applying this operator, each qubit would be measured in the state $1$ with probability $p_n^{\rm fail}$. A circuit for this consists of an $R_y(\theta_n)$ gate\footnote{We use the notation of Qiskit~\cite{qiskit} for the $R_y(\theta)$ gates.} on each qubit as shown in figure~\ref{circuit init}. A probability of $p_n^{\rm fail}$ corresponds to the angle
\begin{equation}
\theta_n=2\,{\rm arcsin}\left(\sqrt{p_n^{\rm fail}}\right).
\end{equation}

The operator $U_{\rm time}$ acts on two consecutive registers. The first register corresponds to time step $t$ and the second register corresponds to time step $t+1$. We use controlled operations to change the qubits corresponding to the nodes.

The operator $U_{\rm time}$ can be constructed as follows:
\begin{itemize}
\item If the qubit $q_n(t)$ corresponding to node $n$ is in state $1$ in time step $t$, then we apply the operation $R_y(\theta)$ with
\begin{equation}
\theta=2\,{\rm arcsin}\left(\sqrt{1-p_n^{\rm recover}}\right)-2\,{\rm arcsin}\left(\sqrt{p_n^{\rm fail}}\right)
\end{equation}
to the corresponding node in time step $t+1$. This can be done with an $1$-controlled $R_y$ operation, i.e. the operation is performed if the control qubit is in state $1$.  After this operation, if the qubit for node $n$ is in state $1$ in time step $t$ then the probability to measure it in state $0$ is $p_n^{\rm recover}$ for a given configurations of the previous time steps as expected from the classical model. Note that the second part of the angle reverts the rotation of a qubit that is performed by $U_{\rm init}$.

\item If a node $n$ is in state $0$ in time step $t$, then we apply several controlled operations $R_y(\theta)$ on qubit $q_n(t+1)$. We define the set
\begin{equation}
M=\{ m \in N : p_{m,n}^{\rm trigger}>0 \}
\end{equation}
of nodes with non-zero probability to trigger node $n$ and we consider all possible configurations $c\in\{0,1\}^{|M|}$
of the nodes in $M$. For a configuration we calculate the probability $p_{\rm off}$ that $n$ stays in state $0$ by the product of the probabilities $1-p_{m,n}^{\rm trigger}$ for all $m \in M$ and $1-p_n^{\rm fail}$, i.e. the node is not triggered by another node and also not intrinsically. Then we add an $R_y(\theta)$ gate on qubit $q_n(t)$ with angle
\begin{equation}
\theta=2\,{\rm arcsin}(\sqrt{1-p_{\rm off}})-2\,{\rm arcsin}(\sqrt{p_n^{\rm fail}})
\end{equation}
and the operation is controlled by the state $0$ or $1$ of qubit $q_m(t-1)$ depending on the state of the configuration $c$.
After this operation, if the qubit for node $n$ is in state $0$ in time step $t$ then the probability to measure it in state $1$ is $1-p_{\rm off}$ for the given configurations of the previous time steps as expected from the classical model.
\end{itemize}

The construction that is outlined above can be optimized in several places, e.g. the separation between $U_{\rm init}$ and $U_{\rm time}$, which makes the description of the construction simpler, is not necessary and introduces controlled operations in $U_{\rm time}$ that revert operations of $U_{\rm init}$.

\subsection{Quantum circuit for example}
\begin{figure}
\centering
\begin{quantikz}
\lstick{$q_1(0)$}\qw &\gate{U_1}&\ctrl{2}   &\octrl{1} &\qw       & \ctrl{1}  & \qw        & \qw        & \qw        & \qw        & \qw\\
\lstick{$q_2(0)$}\qw &\gate{U_2}&\qw        &\ctrl{1}  &\ctrl{2}  & \octrl{2} & \qw        & \qw        & \qw        & \qw        & \qw\\
\lstick{$q_1(1)$}\qw &\gate{U_1}&\gate{U_3} &\gate{U_4}&\qw       & \qw       & \ctrl{2}   & \octrl{1}  & \qw        & \ctrl{1}   & \qw\\
\lstick{$q_2(1)$}\qw &\gate{U_2}&\qw        &\qw       &\gate{U_5}& \gate{U_6}& \qw        & \ctrl{1}   & \ctrl{2}   & \octrl{2}  & \qw\\
\lstick{$q_1(2)$}\qw &\gate{U_1}&\qw        &\qw       &\qw       & \qw       & \gate{U_3} & \gate{U_4} & \qw        & \qw        & \qw\\
\lstick{$q_2(2)$}\qw &\gate{U_2}&\qw        &\qw       &\qw       & \qw       & \qw        & \qw        & \gate{U_5} & \gate{U_6} & \qw
\end{quantikz}
\caption{The quantum circuit for the network in section~\ref{subsection example}. We use the unitaries $U_1=R_y(0.927)$, $U_2=R_y(1.982)$, $U_3=R_y(1.055)$, $U_4=R_y(1.391)$, $U_5=R_y(-1.055)$ and $U_6=R_y(0.135)$. The operators $U_{\rm init}$ are moved to the beginning of the circuit to make the drawing of the circuit more compact.}
\label{circuit example}
\end{figure}
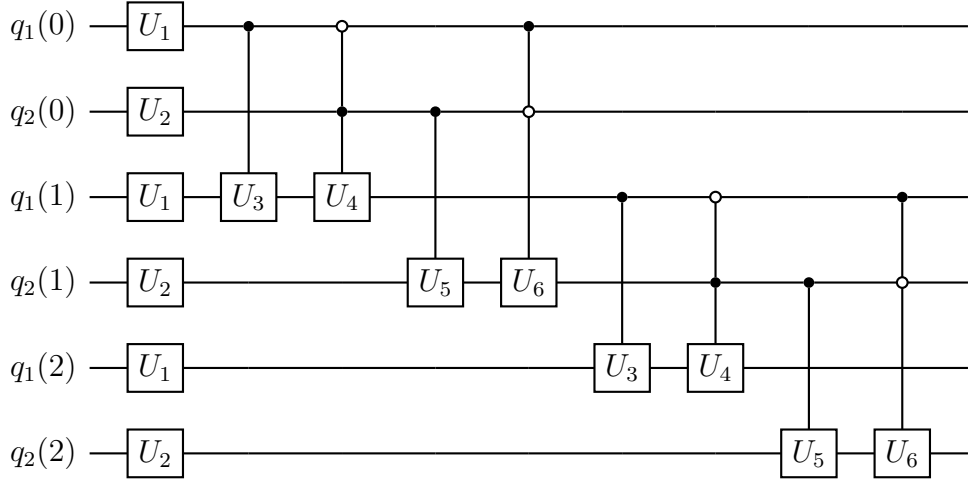

\begin{figure}
\centering
\includegraphics[width=0.8\textwidth]{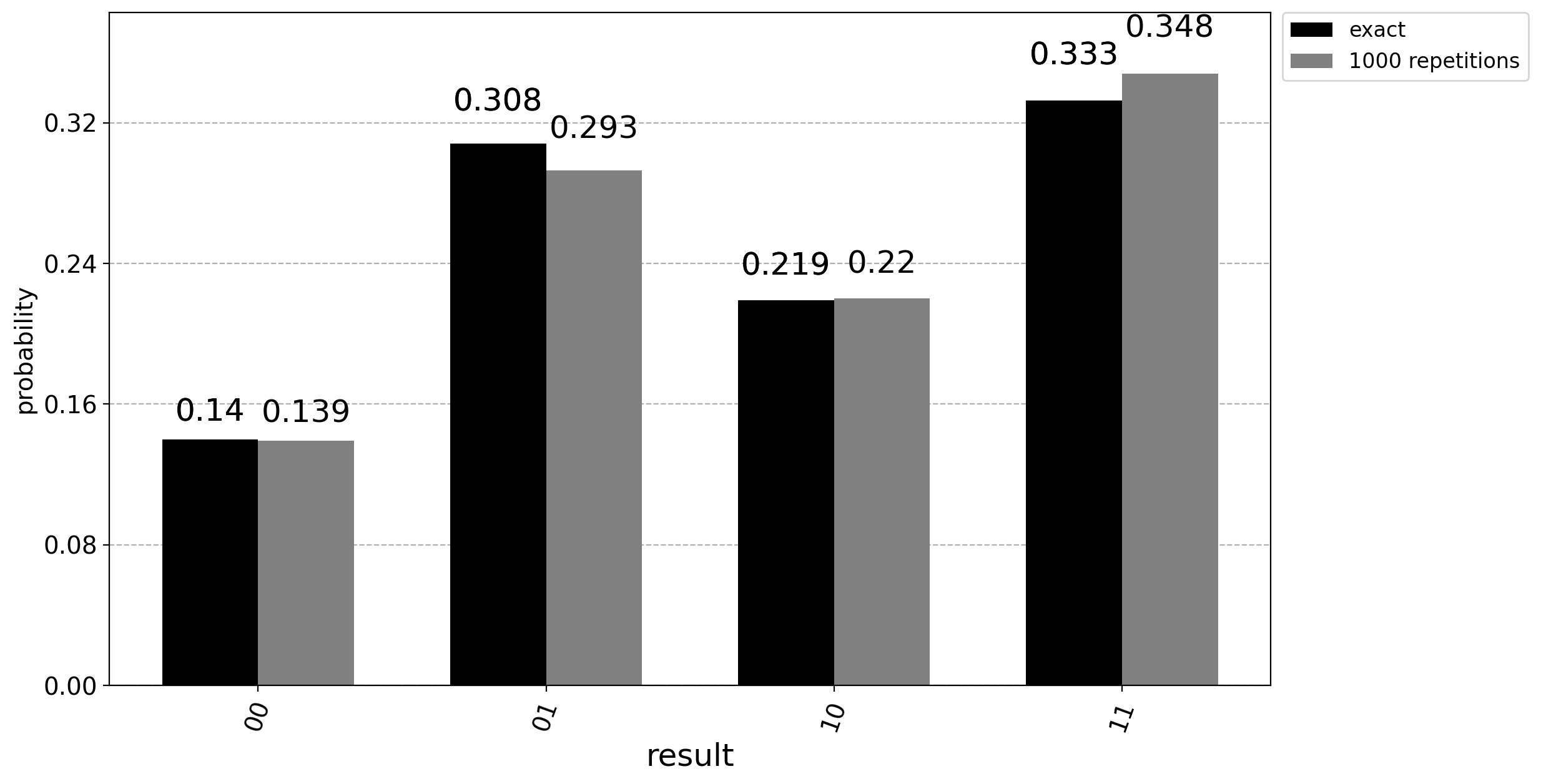}
\caption{Result of the quantum circuit of figure~\ref{circuit example} on an error-free simulator  when we measure the qubits $q_1(2)$ and $q_2(2)$. The probabilities of the measurement results correspond to the probabilities of the configuration of the model in time step $2$.}
\label{fig:QC1Company1TimeStep}
\end{figure}

For the example of section~\ref{subsection example}, the operator $U_{\rm time}$ is shown in figure~\ref{circuit time}:
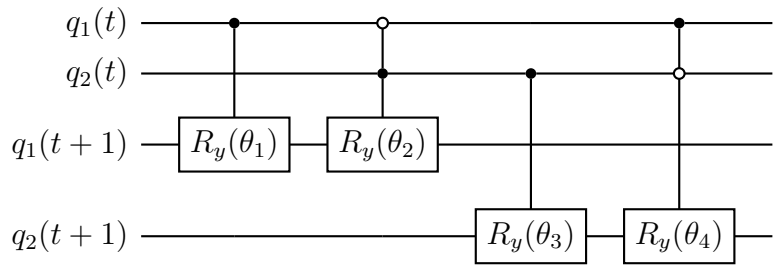
\begin{figure}[b]
\centering
\begin{quantikz}
\lstick{$q_1(t)$}\qw &\ctrl{2} &\octrl{1}  &\qw      & \ctrl{1} &\qw\\
\lstick{$q_2(t)$}\qw &\qw      &\ctrl{1}   &\ctrl{2} & \octrl{2} &\qw\\
\lstick{$q_1(t+1)$}\qw &\gate{R_y(\theta_1)} &\gate{R_y(\theta_2)}        &\qw      & \qw &\qw\\
\lstick{$q_2(t+1)$}\qw &\qw      &\qw        &\gate{R_y(\theta_3)}      & \gate{R_y(\theta_4)}&\qw
\end{quantikz}
\caption{The operator $U_{\rm time}$ for the example from section~\ref{subsection example}. Here, we have the parameters $\theta_1=1.055$, $\theta_2=1.391$, $\theta_3=-1.055$ and $\theta_4=0.135$.}
\label{circuit time}
\end{figure}
The first gate corresponds to the recovery probability of the first node and the angle is determined by
\begin{eqnarray}
&\phantom{=}&2\,{\rm arcsin}\left(\sqrt{1-p_1^{\rm recover}}\right)
-2\,{\rm arcsin}\left(\sqrt{p_1^{\rm fail}}\right)\\
&=&2\,{\rm arcsin}\left(\sqrt{0.7}\right)
-2\,{\rm arcsin}\left(\sqrt{0.2}\right)=1.055.
\end{eqnarray}
The second controlled gate with the angle $\theta_2$ corresponds to the first node that is in state $0$ in time step t and is triggered in time step $t+1$ by the second node when its state is $1$. The probability that node $1$ is not triggered is 
\begin{equation}
(1-p_1^{\rm fail})(1-p_{2,1}^{\rm trigger})=(1-0.2)(1-0.8)=0.16.
\end{equation}
Therefore, the probability that node $1$ is triggered is $0.84$ and this leads to the angle
\begin{equation}
\theta_2=2\, {\rm arcsin}(\sqrt{0.84})-2\,{\rm arcsin}\left(\sqrt{0.2}\right)=1.391.
\end{equation}
The values for $\theta_3$ and $\theta_4$ can be calculated by replacing node $1$ with node $2$ and vice versa.
The full circuit for the example is shown in figure~\ref{circuit example}.
When we measure the last two qubits of the circuit then we obtain the results that are shown
in figure~\ref{fig:QC1Company1TimeStep}.
The probabilities from the measurements and the probabilities of the classical calculations are the same up to the deviations that we expect, because we only performed 1000 repetitions, i.e. we executed the circuit 1000 times on an error-free simulator. With an increasing number of repetitions, the probabilities from the measurements would get closer to the exact values.

If we run the circuit many times and calculate the proportions of the result configurations then we are not more efficient than classical Monte Carlo methods in terms of model evaluations. However, the quantum circuit allows us to use QAE or low-depth QAE methods to evaluate the probability of a configuration with less evaluations of the model on a quantum computer. This is described in the following section.

\section{Quantum amplitude estimation}\label{section QAE}


We can use quantum amplitude estimation (QAE, see~\cite{QAE}) to measure the configuration probabilities of the network time evolution.
On error free hardware, this would lead to a quadratic speedup compared to Monte Carlo simulations in the number of model evaluations. In the following, we consider the standard version of QAE and a low-depth version of it for the example with 2 nodes and 3 time steps that is introduced in section~\ref{subsection example}. The low-depth version is more suited for Noisy Intermediate-Scale Quantum (NISQ) devices as it does not use controlled versions of the Grover operators.

\subsection{Construction of the Grover operator}\label{section Grover operator}
The standard version of the quantum amplitude estimation is a phase estimation of the eigenvalues of a Grover operator. In our case, the network and its time evolution is given by a unitary $U$ that produces the probabilities of configurations after a number of time steps. Based on this operator, we can construct the Grover operator by $G=-U S_0 U^\dagger S_\chi$ where $S_0$ applies the phase $-1$ to the state $|0\ldots 0\rangle$ and $S_\chi$ applies the phase $-1$ to all states that we search. For instance, if we want to find out the probability of a configuration $c \in \{0,1\}^k$ after $t$ time steps, then we mark all states $|c(1)\ldots c(t)\rangle$ with $c(t)=c$ with the phase $-1$. This phase operation acts only on the register corresponding to time step $t$.

For our example from section~\ref{subsection example}, we have 6 qubits for 2 nodes and 3 time steps after the initial configuration. We consider the Grover operators for all 4 possible configurations $00$, $01$, $10$ and $11$ after the last time step.
The eigenvalues of the Grover operators\footnote{We used the function \texttt{numpy.linalg.eig} to calculate the eigenvector from the unitary, which we obtained with the unitary simulator of quantum circuits from Qiskit.} besides $+1$ and $-1$ are given in table~\ref{table eigenvalue}.
We can write
\begin{equation}
\lambda_\pm = {\rm cos}(\theta) \pm i \cdot{\rm sin}(\theta)
\end{equation}
with the values in table~\ref{table eigenvalue} and we obtain the corresponding probabilities
\begin{equation}
p_c(t)={\rm sin}^2(\theta/2)\,.
\end{equation}
We see that for each configuration $c$ the non-trivial eigenvalues of the Grover operator lead to the correct probability of the configuration.
\begin{table}
\centering
\caption{The eigenvalues of the Grover operators corresponding to the 4 possible configurations 00, 01, 10 and 11 for time step $t=3$.}
\vspace{0.5cm}
\begin{tabular}{c|c|c|c}
c & $\lambda_\pm$ & $\theta$ &$p_c(3)$\\
\hline
00&$0.720 \pm 0.694 i$ & $0.767$&$0.140$\\
01&$0.384 \pm 0.923 i$ & $1.177$&$0.308$\\
10&$0.562 \pm 0.827 i$ & $0.975$&$0.220$\\
11&$0.335 \pm 0.942 i$ & $1.230$&$0.333$
\end{tabular}
\label{table eigenvalue}
\end{table}

\vspace{-0.3cm}
\subsection{Standard quantum amplitude estimation}\label{section standard QAE}
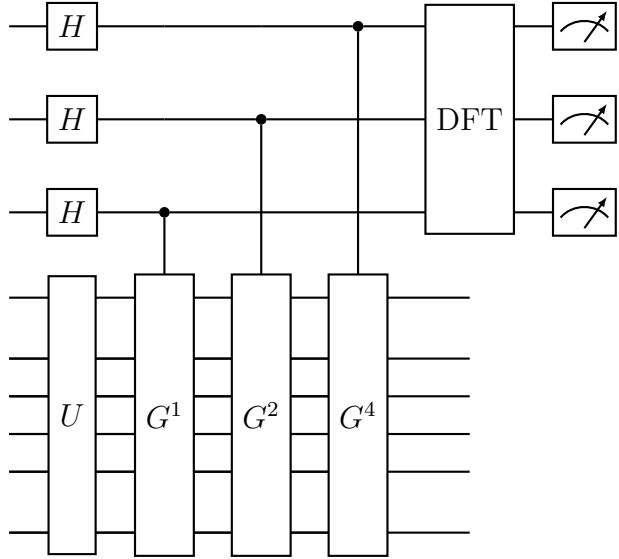
\begin{figure}[h]
\centering
\begin{quantikz}
\qw &\gate{H}    &\qw           &\qw           &\ctrl{3}     &\gate[3]{\rm DFT} &\meter{} & \\
\qw &\gate{H}    &\qw           &\ctrl{2}      &\qw          &                  &\meter{} & \\
\qw &\gate{H}    &\ctrl{1}      &\qw           &\qw          &                  &\meter{} & \\
\qw &\gate[6]{U} &\gate[6]{G^1} &\gate[6]{G^2} &\gate[6]{G^4}&\qw               &      & \\
\qw &\qw         &\qw           &\qw           &\qw          &\qw               &      & \\
\qw &\qw         &\qw           &\qw           &\qw          &\qw               &      & \\
\qw &\qw         &\qw           &\qw           &\qw          &\qw               &      & \\
\qw &\qw         &\qw           &\qw           &\qw          &\qw               &      & \\
\qw &\qw         &\qw           &\qw           &\qw          &\qw               &      &
\end{quantikz}
\caption{The quantum circuit for a standard QAE with a resolution of $3$ qubits. Here, the unitary $U$ is the model of the network and $G$ is the corresponding Grover operator that marks all states depending on the configuration, which we want to analyze.}
\label{circuit standard qae}
\end{figure}

We can use the standard version of QAE to estimate the probability of the
states that are marked by the operator $S_\chi$.
The circuit for an example for the standard version of QAE is given in figure~\ref{circuit standard qae}.

The unitary operation $U$ is the model and it is used to initialize the registers, on which the controlled Grover operators $G^\ell$ act.
The measurements after the Fourier transform lead to a binary encoding of the probability, which we want to find. For an increasing number of qubits we obtain
an increasingly better approximation of the result. For a resolution of three qubits as in figure~\ref{circuit standard qae} we obtain eight possible binary results along with the angles $\theta$ and the corresponding probabilities ${\rm sin}^2(\theta/2)$ that are shown in 
table~\ref{table bin probabilities}.


\begin{table}[h!]
\centering
\caption{The binary results with their corresponding angles $\theta$ and probabilities
for a QAE with 3 bits resolution.}
\vspace{0.5cm}
\begin{tabular}{c|c|c|c|c|c|c|c|c}
res & 000 & 001 & 010 & 011 & 100 & 101 & 110 & 111 \\
\hline
$\theta$ & 0.0 & 0.785 & 1.571 & 2.356 & 3.142 & 3.927 & 4.712 & 5.498 \\
prob & 0.0 & 0.146 & 0.500 & 0.854 & 1.0 & 0.854 & 0.500 & 0.146
\end{tabular}
\label{table bin probabilities}
\end{table}

\begin{figure}[h]
\centering
\includegraphics[width=1.0\textwidth]{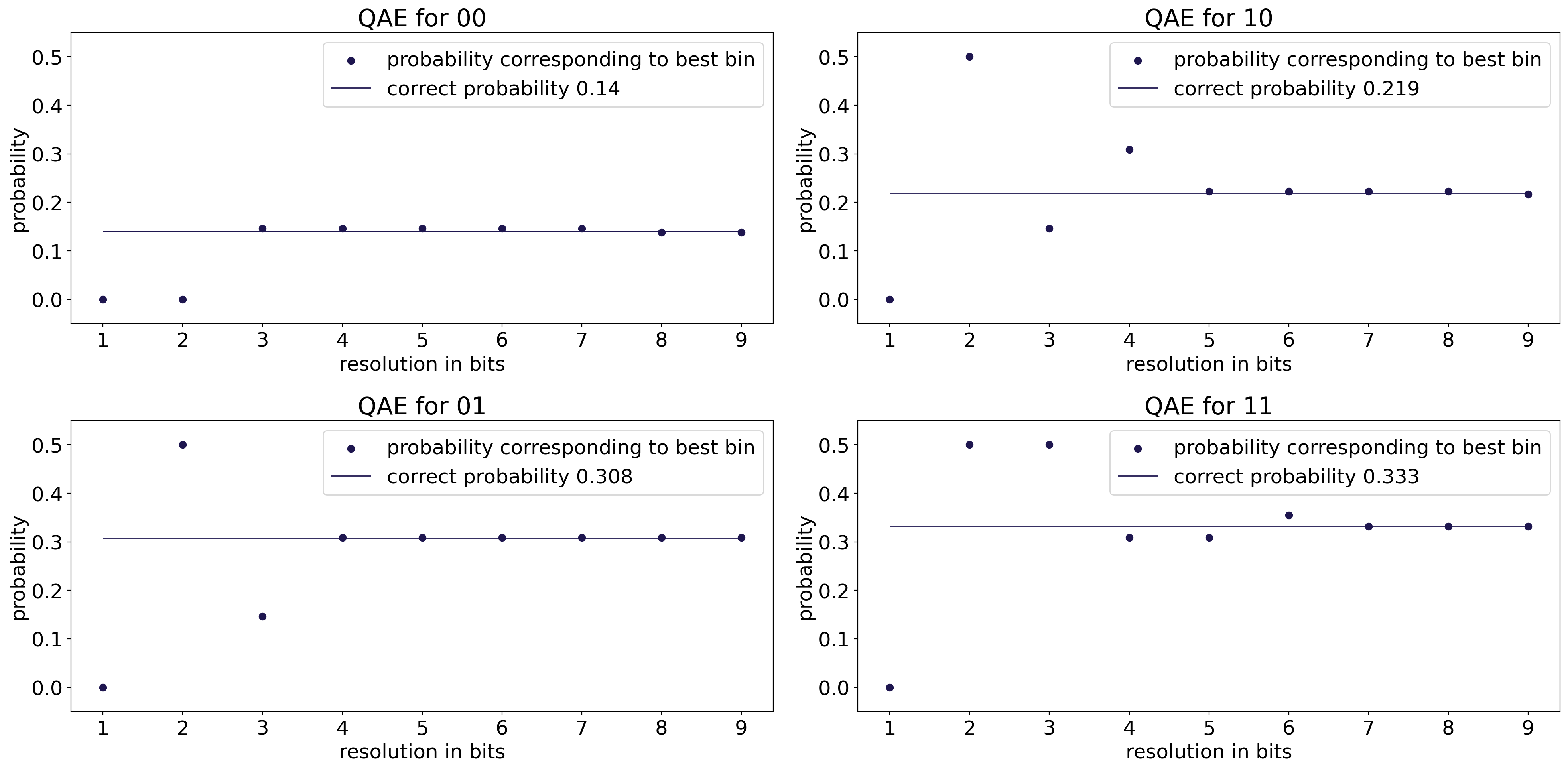}
\caption{The probabilities obtained by QAE depending on the number of qubits of the precision. }
\label{result QAE}
\end{figure}

In figure~\ref{result QAE}, we show the results of a QAE depending on the number of qubits that are used for the resolution of the output.
The resulting probabilities are the values that correspond to the binary result with the highest number of counts. We see that with an increasing number of qubits the results are getting closer to the exact value.

The disadvantage of this method for amplitude estimation is that we need controlled operators, which are the powers of Grover operators. Without finding an exploitable structure of a Grover operator, we can only apply a Grover operator $\ell$ times for performing the operator $G^\ell$. This means that for a QAE with $b$ qubits of precision we have in total $2^b-1$ controlled Grover operators in the circuit. Each Grover operator contains the model and its inverse along with additional operators to mark states with the phase $-1$. The model typically contains operations on qubits that are controlled by several control qubits and the controlled version of the Grover operator adds an additional qubit to this. As a consequence, the implementation of a QAE for even very simple models and for very few time steps on currently available hardware appears to be impossible. 

For small circuits, the replacement of operators, which have several control qubits, with more elementary gates adds a significant overhead to a circuit. For circuits with many control qubits, the problem is less severe as there are techniques to reduce gates with several control qubits to more elementary gates with a linear overhead~\cite{barenco}. 

An alternative to the QAE is given by low-depth versions of QAE. These methods try to avoid the additional control qubit for the Grover operators and they trade this reduction of quantum complexity for a higher post-processing complexity. One of such methods is described and applied in the following section.

\subsection{Low-depth quantum amplitude estimation}\label{subsection low-depth}
In this section we use a method for amplitude estimation that is inspired by~\cite{QAEwoPhase} to reduce the necessary hardware requirements. The main advantage of this construction is that controlled versions of the Grover operators are not needed. Besides this, we also do not need the additional qubits for the controlled operations and the Fourier transform. We execute several Grover operators directly on a qubit register after an initialization with the model and measure the number of good results. The fact that we do not need controlled Grover operators makes it reasonable to run such quantum circuits on current hardware.

\begin{figure}
\centering
\begin{quantikz}
&\qwbundle{q} &\gate{U}    &\gate{G} &\gate{G}  &\gate{G} &\meter{} & \\
\end{quantikz}
\caption{Circuit for low-depth QAE for a model $U$ and a Grover operator $G$ on a register with $q$ qubits. The number of Grover operators depends on the chosen specific low-depth method. We use the measurement results to determine  the proportion of states, which are marked by $S_\chi$ of the Grover operator.}
\label{circuit low-depth qae}
\end{figure}
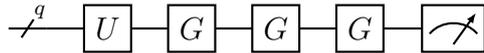

A simple schematics for a low-depth QAE is shown in figure~\ref{circuit low-depth qae}.
As an example, we calculate the probabilities $p_c(t)$ of section~\ref{subsection example} of all four possible configurations after 3 time steps. For this, we use the 4 different Grover operators from section~\ref{section Grover operator} and we run the series of experiments for each case on an error-free simulator. We execute the Grover operators $G^\ell$ for all $\ell \in \{0,1,\ldots,8\}$ after the initialization of the register with $U$ and for each such power we run the circuit 30 times\footnote{Note that this simple choice of the powers of the Grover operators and the constant number of repetitions is chosen to give a good overview of the method.}. In table~\ref{table results simulator 2 nodes} we list the obtained counts $m_\ell$ for all configurations.

\begin{table}[h]
\centering
\caption{The counts for the low-depth QAE with $0$ to $8$ Grover operators after the initialization with $U$ for all possible configurations of the example with 2 nodes and 3 time steps of section~\ref{subsection example} on a noise-free simulator.}
\vspace{0.5cm}
\begin{tabular}{c|ccccccccc|c|c}
c & $m_0$ & $m_1$ & $m_2$ & $m_3$ & $m_4$ &$m_5$ & $m_6$ & $m_7$ & $m_8$ & $\theta$ &$p_c(3)$\\
\hline
00 & 5& 29& 26&  5&  4& 21& 27&  6&  6& 0.390& 0.144\\
01 & 8& 27&  2& 22& 19&  0& 30& 10&  7& 0.588& 0.308\\
10 & 6& 29& 10&  1& 25& 18&  0& 24& 25& 0.487& 0.219\\
11 & 7& 28&  0& 22& 16&  7& 30&  2& 23& 0.612& 0.330
\end{tabular}
\label{table results simulator 2 nodes}
\end{table}

\begin{figure}[h]
\centering
\includegraphics[width=0.8\textwidth]{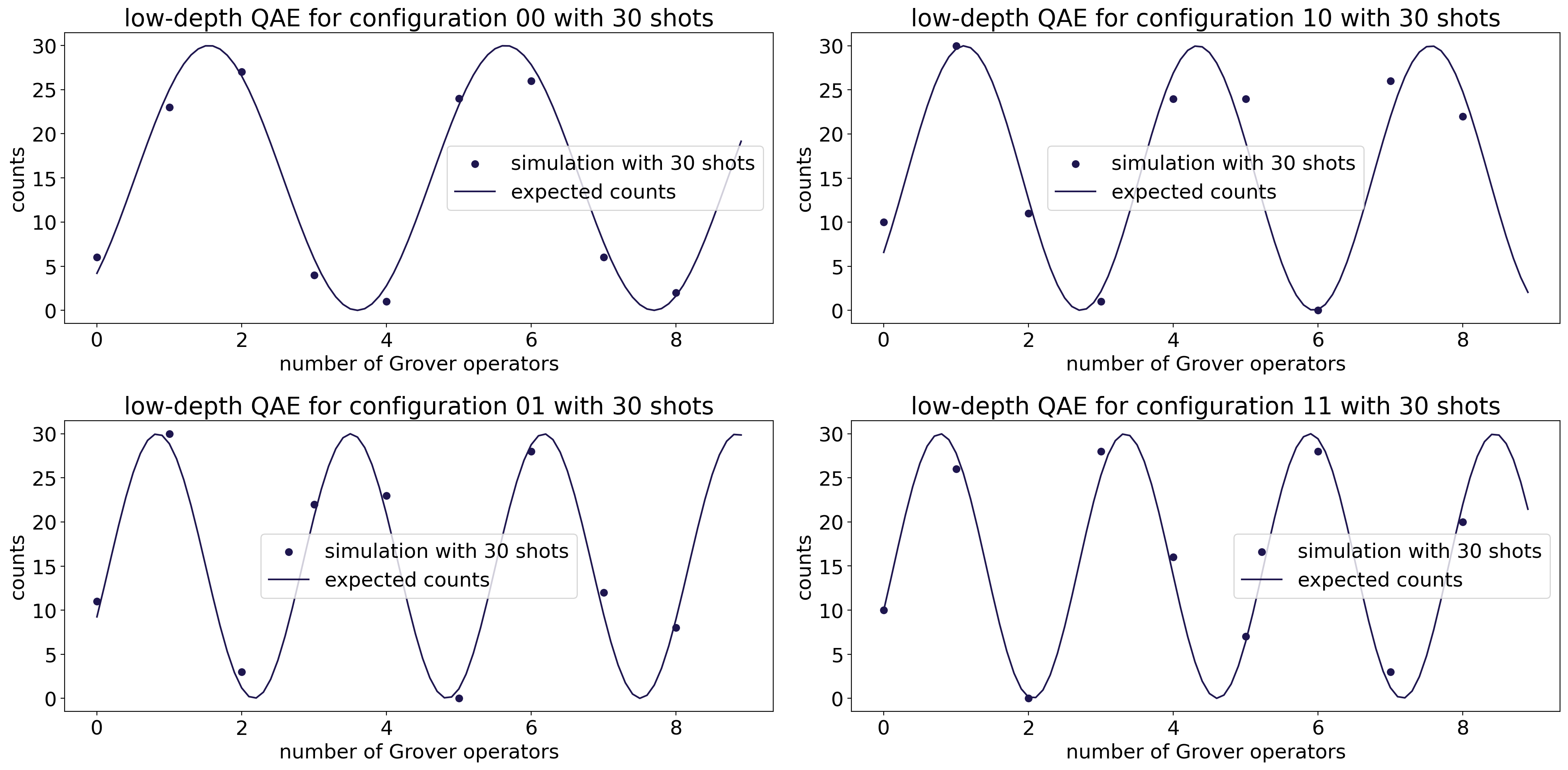}
\caption{The results of the low-depth QAE with 30 repetitions for $\ell$ Grover oracles with $\ell \in\{ 0, \ldots, 8 \}$ on an error-free simulator. We consider the Grover oracles for the configurations $00$, $01$, $10$ and $11$ separately. The solid curves are ${\rm sin}^2((2\ell+1)\theta/2)$ for the correct values of $\theta$ from the exact classical calculation. We show the correct values for real-valued $\ell$ instead of integer values to show the origin of the values more clearly.}
\label{result low depth QAE}
\end{figure}

For instance, the value of $m_2$ denotes the number of  measurements 
of marked configurations out of the 30 repetitions when we apply two Grover 
operators after the initialization. 
For an angle $\theta$, which we want to determine, and the power $\ell$ of a Grover operator we obtain the corresponding probability 
\begin{equation}\label{equation sine}
{\rm sin}^2((2\ell+1)\theta/2)
\end{equation}
for measuring the good result. Therefore, for any angle $\theta$ we can compare the deviation of the expected values $m_\ell$ from the measured results. This deviation can be used to perform a gradient descent search for the angle $\theta$. The angles and probabilities in table~\ref{table results simulator 2 nodes} were obtained by this method\footnote{The corresponding functions for the gradient descent method are implemented in the Pygrnd module {\tt pygrnd.qc.lowDepthQAEgradientDescent}.}. In figure~\ref{result low depth QAE} we show the measured points together with the probabilities based on the correct angle $\theta$, which we derived from the exact classical calculation. The results show that the method works for an error-free simulation of the quantum circuits for low-depth QAE.

Note that the function of equation~(\ref{equation sine}) has two solutions for each probability value, e.g. we obtain the probability 0.3 for $\theta=1.159$ and $\theta=1.982$. If $\theta$ is an angle, then
$2\pi-\theta$ is the other angle that leads to the same probability. In the following, we always take the smaller angle of both.

\subsection{Low-depth quantum amplitude estimation with noise}\label{subsection low depth noise}
We can try to extract a useful result from the output of a noisy quantum computer by using our knowledge of the output functions. An example for such an approach, which differs from ours, was presented in~\cite{noiseAware}. The authors derived the shape of the distributions of the measurement results for low-depth QAE on a noisy machine by assuming that each rotation angle of a Grover picks up a normally distributed error. Other results for the output of low-depth QAE on noisy hardware were presented in~\cite{JoS2022} for parallel versions of QAE. An approach to noise that is similar to the one in this section was discussed in~\cite{LDQAE} for three different low-depth QAE algorithms. Two of those algorithms were tested
on quantum hardware in~\cite{LDQAEhardware}.

Here, we want to model the number of results, which are marked by $S_\chi$, that we measure at the end of a noisy low-depth QAE circuit. We start from the assumption, that even a single error in a Grover operator will cause the output of the Grover operator to be essentially random. Therefore, when one or more errors occur, we expect that the ratio of marked states, which we measure at the end of the circuit, is the fraction $f$ of states, which are marked by $S_\chi$, among all possible states.

Furthermore, we assume that for circuits of the low-depth QAE as in figure~\ref{circuit low-depth qae} with only the $M$ operator and no Grover operators the effect of noise can be neglected, i.e. we expect the results from equation~(\ref{equation sine}).
We also know that the probability that at least one error occurs during the execution of a circuit grows exponentially with the gate count. Therefore, we add an exponential decay factor to the expected  output probability of equation~(\ref{equation sine}). For a circuit QAE with angle $\theta$ and $\ell$ Grover operators we denote by $r(\theta, \ell, a, f)$ the expected probability for measuring a marked result, where $a$ approximates the probability of one Grover operator to incur an error when executed. This probability is given by the following equation:
\begin{equation}\label{errormodel}
r(\theta, \ell, a, f)= e^{-a \ell} {\rm sin}^2((2\ell+1) \theta/2) + (1 - e^{-a \ell}) f
\end{equation}
We can use equation~(\ref{errormodel}) to fit the output of a quantum computer and extract the angle $\theta$ and therefore the probability to find good states. We apply this method to an example in section~\ref{section hw}.

The expected benefits of using this model of a dampened oscillation over the naive approach of fitting the results to a pure sine wave are two-fold: First, the fitter should have an advantage due to the fact that the data points are generally closer to the fit, due to the dampening. Second, the dampening changes the wave length of the oscillation and this introduces an error in the naive fitting. The wavelength we find is too long and the corresponding probabilities are too low. For example, this effect can be seen in figure~\ref{result 1 node error model hardware}). With the error model, this problem should not arise.

\subsection{Implications of the noise model}
It is instructive to think of the simple error model of section~\ref{subsection low depth noise} as the sum of two different error models. The first one determines if one or more errors do occur in a circuit. This is modeled by the $e^{-a\ell}$ terms in equation~(\ref{errormodel}). The second model determines the output in case there is an error. If we assume that an error makes the output completely random, which is the simplest possible approximation, then the factor $f$ in equation~(\ref{errormodel}) is the expectation value of our second error model.

When we explicitly model these errors as a sum over Bernoulli random variable with expected value $f$, then we see that the factor $f$ leads to the standard deviation
\begin{equation}
\sigma = \sqrt{\frac{f (1-f)}{N}}\,.
\end{equation}
We want to determine the number $\ell$ of consecutive Grover operators, which we can expect to execute and which allows us to still measure a useful signal at the end of the circuit. This number depends on the noise level and we can simply compare the amplitude ${\rm max}_\ell e^{-a\ell}$ of the signal in equation~(\ref{errormodel}) with the mean absolute deviation ${\overline \sigma}$, which is 
\begin{equation}
{\overline \sigma}=\sqrt{\frac{2}{\pi}}\sigma\,,
\end{equation}
where $N$ denotes the number of repetitions, which we perform for a circuit.
If the mean absolute deviation is bigger than the amplitude of the signal, then we cannot hope to extract the sine wave from our measurements anymore. This means that if the inequality
\begin{equation}
\sqrt{\frac{2}{\pi}} \cdot \sqrt{\frac{f (1-f)}{N}} < e^{-a \ell}
\end{equation}
does not hold, then we cannot expect to measure useful results. 
Solving this for $N$ tells us how many repetitions we need to find a useful signal for a given noise level $a$ and number of Grover operators $\ell$.
If we solve it for $\ell$, we find how many Grover operators we can have in a low-depth QAE for a given noise level $a$ and number of repetitions $N$.

We can now try and use those insights to obtain optimal results on actual quantum hardware. This means building a protocol which, after the initial circuit with just the model evaluation, determines the optimal number of Grover operators and repetitions for the next experiment, based on the results known so far. We can also determine an estimation for an upper limit for the circuit with the largest number of Grover operators already after the second experiment. Some results related to such ideas have been worked out in~\cite{LDQAE}.

\subsection{Evaluation of results from simulator with noise model}\label{section hw}
The example of section~\ref{subsection example} with $2$ nodes and $3$ time steps is too complex  for current hardware due to the need of several gates in each Grover operator that are controlled by more than one qubit. For most of the currently available hardware types, these operators have to be decomposed into single-qubit operators, which are uncontrolled or which are controlled by a single qubit~\cite{barenco}. This decomposition introduces an overhead, which makes the evaluation of the low-depth QAE infeasible.

\begin{figure}
\centering
\begin{quantikz}
\lstick{$q_1(1)$} & \gate{R_y(1.159)} & \ctrl{1} & \qw             &\qw       &\qw             &\qw \\
\lstick{$q_1(2)$} & \gate{R_y(1.159)} & \gate{R_y(1.339)}           & \ctrl{1} &\qw  &\qw           &\qw \\
\lstick{$q_1(3)$} & \gate{R_y(1.159)} & \qw      & \gate{R_y(1.339)}&\qw       &\ctrl{1}        &\qw \\
\lstick{$q_1(4)$} & \gate{R_y(1.159)} & \qw      & \qw             &\qw       &\gate{R_y(1.339)}&\qw
\end{quantikz}
\caption{Circuit for the network model with 1 node and 3 time steps after the initialization. We are interested in the probability that the qubit $q_1(4)$ is in state 1.}
\label{circuit low-depth noise sim}
\end{figure}
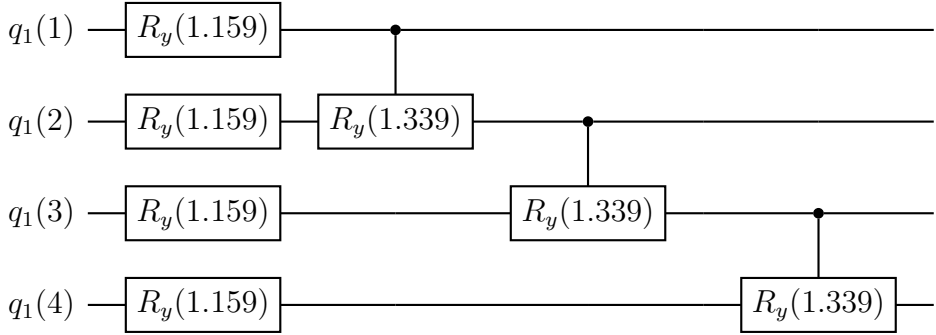

We analyze the behavior of the low-depth version of QAE as described in section~\ref{subsection low depth noise}. For this,  we consider a simpler network and we run it on the AQT simulator~\cite{aqt-qiskit}, which uses a realistic noise model for  ion-trap quantum computers~\cite{aqt-noise-model}. The model has only one node and we consider up to $3$ time steps after the initial step. The node has the probabilities $p_1^{\rm fail}=0.3$ and $p_1^{\rm revover}=0.1$. The probability that the node is in state 1 after $t$ time steps is shown in table~\ref{probs 1 node}.
\begin{table}
\centering
\caption{The probabilities that we have state 1 after $t$ time steps for the example with one node from the exact classical calculation.}
\vspace{0.5cm}
\begin{tabular}{c|cccc}
t & 1 & 2 & 3 & 4 \\
\hline
probability & 0.300 & 0.480 & 0.588 & 0.653
\end{tabular}
\label{probs 1 node}
\end{table}
The quantum circuit for the model is shown in figure~\ref{circuit low-depth noise sim}.

We construct the Grover operators for all time steps $t\in \{1,2,3,4\}$ and we run the low-depth version along with the gradient descent search for the parameters $a$ and $f$ and the probabilities as described in section~\ref{subsection low depth noise}.
We use the AQT simulator with noise and we make 2000 repetitions for each operator and each number of Grover operators in the low-depth circuit. The results are shown in table~\ref{table results simulator}.

\begin{figure}[h!]
\centering
\includegraphics[width=1.0\textwidth]{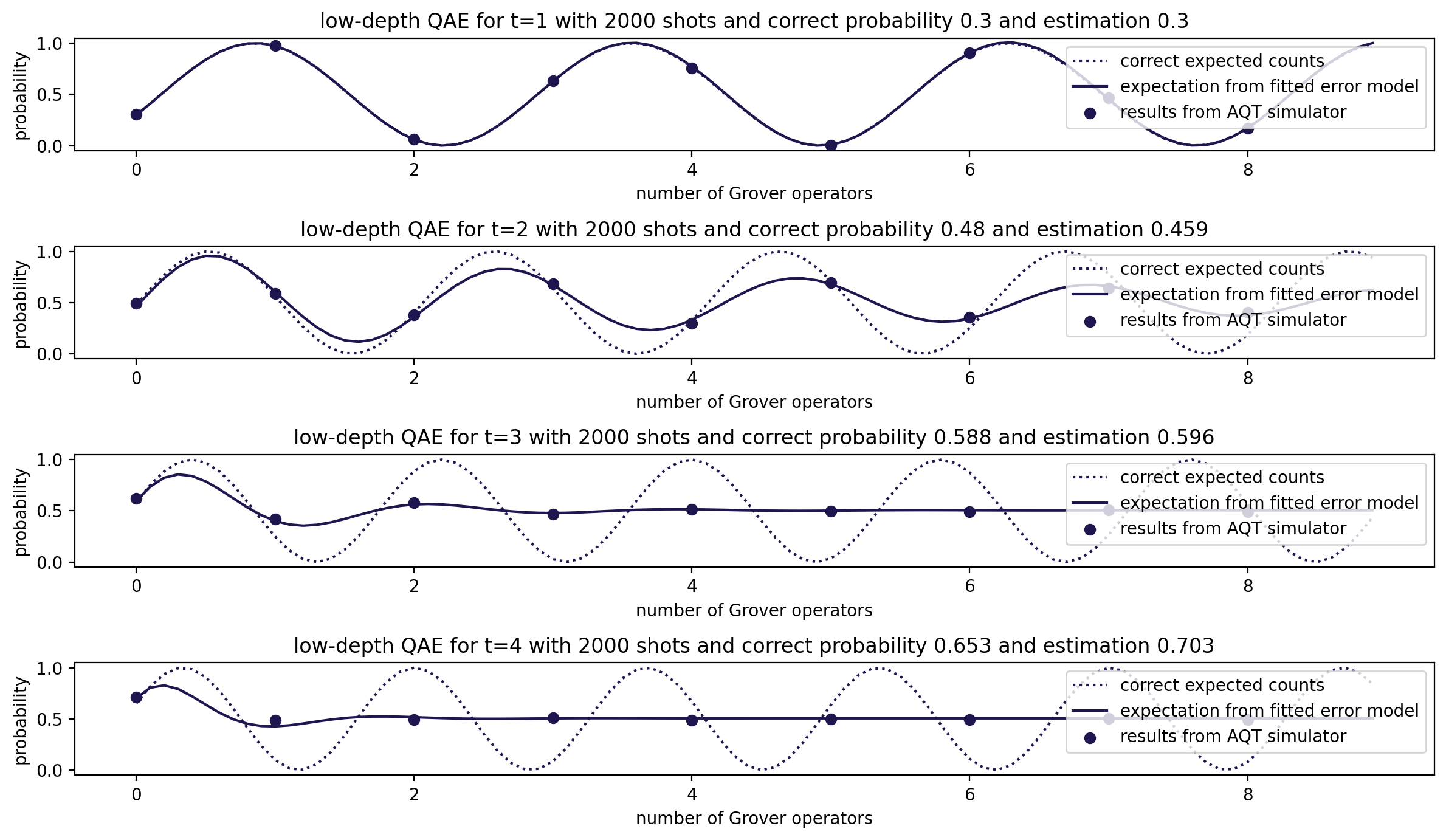}
\caption{Measurement results (simulator with noise model) for the model with $1$ node (filled circles) and expected probabilities to find the state 1 for the values, which are calculated by the exact classical method (dotted curves). The solid curves are the expected probabilities from equation~(\ref{errormodel}) with values $\theta$, $a$ and $f$ after gradient descent optimization. For $t=1$, only one curve can be seen as both are very close.}
\label{result 1 node error model}
\end{figure}

\begin{table}[h]
\caption{The counts for the low-depth QAE with $0$ to $8$ Grover operators after the initialization with $U$ for the example with 1 node and different time steps on the AQT simulator with noise model with 2000 repetitions for each $t$ and $m_\ell$. The values $a$ and $f$ and the probabilities result from the gradient descent when we have the measured values $m_\ell$.}
\vspace{0.5cm}
\begin{tabular}{c|ccccccccc|ccc}
t & $m_0$ & $m_1$ & $m_2$ & $m_3$ & $m_4$ &$m_5$ & $m_6$ & $m_7$ & $m_8$ & a& f &prob\\
\hline
1 &  617& 1958&  121& 1265& 1516&   12& 1817&  928& 344 & -0.001 & 0.003 & 0.300 \\
2 &  981& 1181&  760& 1365&  595& 1384&  717& 1283& 807 & 0.162  & 0.513 & 0.459 \\
3 & 1236&  841& 1153&  931& 1033&  988&  983& 1018& 979 & 0.977  & 0.504 & 0.596 \\
4 & 1423&  969&  991& 1026&  979& 1002&  983& 1012& 986 & 1.813  & 0.506 & 0.703
\end{tabular}
\label{table results simulator}
\end{table}
We see that for one or two time steps the values for the probability are accurate. However, for 3 and 4 time steps the exponential error factor $a$ is much higher and this shows that there is a significant amount of error. This can be explained by the fact that in the circuit with 3 time steps we have the operator $S_0$ that is an $X$ gate that is controlled by two qubits.  After compiling to 1- and 2-qubit gates, this leads to a much longer circuit compared to the circuit for 2 time steps. The measurement results with the values of equation~(\ref{equation sine}) for the correct values without error model, which are calculated with the exact classical method, are shown in figure~\ref{result 1 node error model} along with the values of equation~(\ref{errormodel}) for the values, which we found with gradient descent for the error model.

\subsection{Evaluation of results from hardware}\label{section hardware run}
In addition to the evaluation of a network model on a simulator in section~\ref{section hw}, we also evaluated the model on the AQT ion trap quantum computer PINE. The PINE system was configured to work with a register of 8 qubits for our implementation. We used the Qiskit function {\tt transpile} with optimization level 3 to convert the circuits into native gates for this machine.\footnote{With this optimization level, the circuits for $t=1$ consist of a single uncontrolled gate.} For $t=1$ and $t=2$ we considered up to 8 Grover operators after the initialization and for $t=3$ and $t=4$ we used up to 4 Grover operators. For each number $t$ of time steps and number $\ell$ of Grover operator
we gathered statistics from executing a circuit 8000 times. We used the simple error model from section~\ref{subsection low depth noise} to obtain an estimation of the probabilities.  The counts and the values $a$ and $f$ of the model, which we fitted with gradient descent, and the resulting estimated probabilities to find the node in state 1 after $t$ time steps are shown in table~\ref{table results hardware}.

\begin{figure}[h]
\centering
\includegraphics[width=1.0\textwidth]{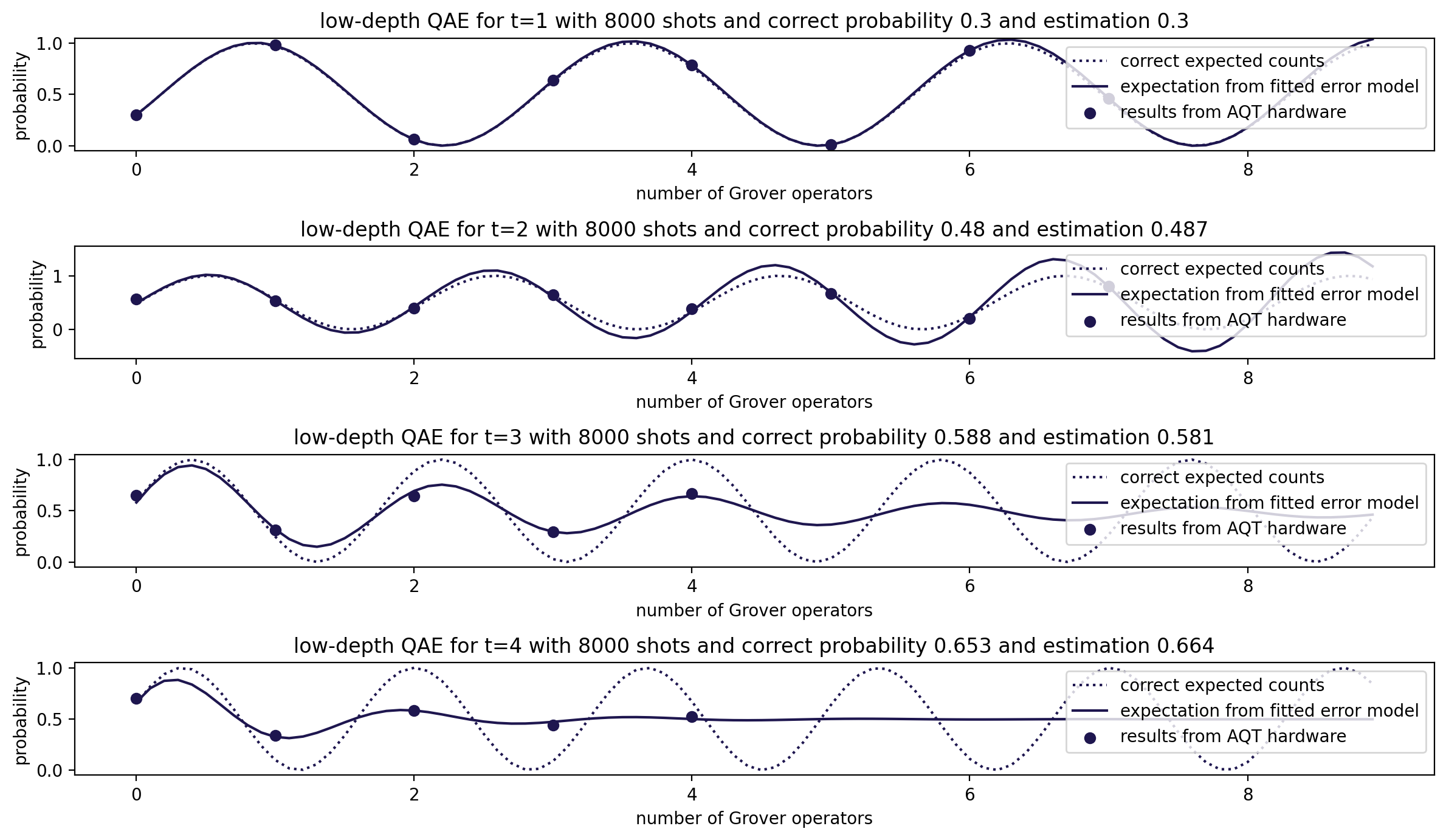}
\caption{Measurement results (on PINE) for the model with $1$ node (filled circles) and expected probabilities to find the node in state 1 for the correct value (dotted curves). The solid curves are the expected probabilities from equation~(\ref{errormodel}) with values $\theta$, $a$ and $f$ after gradient descent optimization. For $t=1$, only one curve can be seen as both are very close. Note that the wavelength of the oscillation of the solid curves for $t=3$ and $t=4$ are shorter, which illustrates the advantage of the error model, see section \ref{subsection low depth noise}. }
\label{result 1 node error model hardware}
\end{figure}

We see that the estimation based on the error model is quite close to the correct values from the exact classical evaluation, which are in the first table in section~\ref{section hw}. The data points, the exact values and the function of the simple error model, which is fitted with gradient descent, are shown in figure~\ref{result 1 node error model hardware}.
Note that for $t=2$, we have negative probabilities\footnote{The assumption that the complexity of the circuits increases with the number of Grover operators does not hold for $t=1$ and $t=2$. The reason for this is that the optimizer is able to reduce the number of gates to a constant number.} from the model after fitting. Despite this 
result, the estimation of the probability is very close.

\begin{table}
\centering
\caption{The counts for the low-depth QAE with $0$ to $8$ Grover operators for $t=1$ and $t=2$ and with $0$ to $4$ operators for $t=3$ and $t=4$ after the initialization with $U$ for the example with 1 node on the AQT system PINE. The values $a$ and $f$ and the probabilities result from the gradient descent when we have the given values $m_\ell$. The values in the column $p_{\rm cor}$ are the results from the exact calculation.}
\vspace{0.5cm}
\begin{tabular}{c|cccccccc|cccc}
t & $m_0$ & $m_1$ & $m_2$ & $m_3$ & $m_4$ &$m_5$ & $m_6$ & $m_7$  & a & f &prob&$p_{\rm cor}$\\
\hline
1 & 2386& 7867&  494& 5107& 6283&   84& 7450& 3689& -0.006 & 0.060 & 0.300 & 0.300\\
2 & 4545& 4275& 3133& 5118& 3110& 5316& 1639& 6446& -0.076 & 0.526 & 0.487 & 0.480\\
3 & 5211& 2505& 5171& 2336& 5353&  -  &  -  &  -  &  0.286 & 0.476 & 0.581 & 0.588\\
4 & 5609& 2695& 4678& 3510& 4171&  -  &  -  &  -  &  0.884 & 0.497 & 0.664 & 0.653
\end{tabular}
\label{table results hardware}
\end{table}

We tried to improve the fitting of the error model to the data by setting $f=0.5$, which corresponds to the proportion of states, which are marked by $S_\chi$, to all states, but this did not improve the estimation.\\
As can be seen from table \ref{table results hardware}, for $t=3$ and $t=4$, our noise model results were within about 1 and 2 percentage points of the exact results. In contrast to this, without noise model we could not even get to within 5 and 14 percentage points (for details see the notebook {\tt probabilisticNetworks.ipynb} in Pygrnd~\cite{pygrnd}). This illustrates the value added by the noise model.

\section{Conclusion and outlook}
We showed how a simple model for the discrete time evolution of networks 
can be formulated as quantum circuits. We used the circuits to apply quantum amplitude estimation and a low-depth version of it to find the probabilities of different network configurations and their time evolution for simple examples. We introduced a simple error model for the expected measurement results on noisy hardware, which can be used to improve the post-processing of the measurement results. We used this approach to analyze a simple example on a simulator with a realistic noise model as well as on real hardware.

It is straightforward to extend the model to more dependencies between the nodes, e.g. a recovered node could lead to the recovery of another node with a certain probability. Another useful extension would be to allow dependencies over several time steps.

It is encouraging to see that already today, simple but non-trivial models can be evaluated not just on simulators, but on real life quantum computing hardware, AQT's PINE machine in this case. The impact of the error model seems worth further investigations, even though significant hardware advances will be required before quantum computing will find practical applications for evaluating models like the one presented here.

\section*{Acknowledgement}
This work is supported by the German Federal Ministry of Education and Research
through project 'Quantencomputer mit gespeicherten Ionen für Anwendungen (ATIQ)', sub-project 'Modellierung von Kreditrisiken mit Quantenalgorithmen' (FKL: 13N10124).

We also acknowledge support by the Austrian Research Promotion Agency (FFG) through grants ELQO (FFG-No 884471), HPQC (FFG-No 897481) and ITAQC (FFG-No 896213), and by the EU Quantum Technology Flagship under the MILLENION Grant Agreement with no. 101114305.

\begin{appendices}
\titleformat{\section}[display]
    {\normalfont\Large\bfseries}{\appendixname\enspace\thesection}{.5em}{}
\section{Pygrnd implementation of the exact evaluation}\label{appendix 1}
In this section, we describe the algorithm of Pygrnd~\cite{pygrnd}, which calculates the exact probabilities with a classical algorithm. This evaluation is implemented in the function {\tt classicalEvaluation} in the module {\tt pygrnd.qc.probabilisticNetworks}.

We can simplify the description of the algorithm by defining the following data structure. 
For a subset $M \subseteq N$ and for a time step $t$, we denote by $C_M(t)$ a data structure, which stores pairs $(b,c) \in \{0,1\}^k \times \{0,1\}^k$ along with their corresponding probabilities after processing the nodes in $M$. A pair can occur more than once with different probability values. Here, the parameter $M$ denotes the structure after all the nodes in $M$ are already processed. The structure $C_{\{\}}(t)$ denotes the data before processing a node in this time step. The configuration $b$ in the pair $(b,c)$ is used to keep track of the history of a configuration during calculations, e.g. if we start with the configuration $00$ and we calculate the probabilities of $00$ and $10$ for the next time step by considering first the probabilities for the first node, then we need to know the original configuration of $10$ because the probabilities of the second node depend only on the configuration of the previous time step. The value $b$ of the pair gives us this information.

The following iterative method makes use of this data structure and the method allows us to calculate the probabilities $p_c(t+1)$ when we have the probabilities $p_c(t)$ for all possible configurations $c\in \{0,1\}^k$:

\begin{itemize}

\item We start with the configuration $c=(0,\ldots,0)$ and assign the probability $p_c(0)=1$ for time step $t=0$ to it. This is the initialization of the procedure. It means that in $C_M(0)$ we have only the element $(c,c)$ with probability 1.

\item For each new time step $t+1$, we iterate over all configurations $(b,c) \in C_N(t)$, i.e. the data structure after processing all nodes in time step $t$. We store the pairs $(c,c)$ along with the probability of $p_{(b,c)}(t)$ in $C_{\{\}}(t+1)$. If during the iteration over all elements of $C_N(t)$ a pair $(c,c)$ is already in $C_{\{\}}(t+1)$ then we add the corresponding probability to the existing value.

\item When the nodes in the current subset $M \subseteq N$ are already processed, then we consider the next node $n \in N$. We start with an empty $C_{M^\prime}(t+1)$ where $M^\prime = M \cup \{n\}$.

\item We iterate through all elements $(b,c) \in C_M(t+1)$. The elements have the form $(b,c)$ and $b$ is the original configuration from the previous time step.

\begin{itemize}
\item If node $n$ is set to $1$ in the value $b$, then we add $(b,c)$ to $C_{M^\prime}(t+1)$ with the probability value $1-p_n^{\rm recover}$ and the value $(b,c^{\prime})$ with probability $p_n^{\rm revocer}$, where $c^\prime=(s_1^\prime(t), \ldots, s_k^\prime(t))$ with $s_n^\prime(t)=0$ and $s_m^\prime(t) = s_m(t)$ for $m\in N\setminus \{n\}$.

\item If node $n$ is set to $0$ in the value $b$, then we calculate the probability $p_{\rm off}$ that the node stays off in time step $t+1$. This is the product of $1-p_n^{\rm fail}$ and the probabilities $1-p_{a,n}^{\rm trigger}$ for all nodes $a$ that are set to $1$ in value $b$. We add $(b,c)$ to $C_{M^\prime}(t+1)$ with probability value $p_{\rm off}$ and $(b,c^\prime)$ with $1-p_{\rm off}$ where $c^\prime=(s_1^\prime(t), \ldots, s_k^\prime(t))$ with $s_n^\prime(t)=1$ and $s_m^\prime(t) = s_m(t)$ for $m\in N\setminus \{n\}$.
\end{itemize}

\item If all entries of $C_M(t+1)$ are processed then we move to the next node. If all nodes are processed for a time step $t$ then we can set $C_{\{\}}(t+1)$ to $C_M(t)$ and start with iterating over the nodes again.

\item Note that for a pair $(b,c)$ of a configuration the element $b$ stays constant through all nodes for a time step. We use $b$ to keep track of the configuration of the nodes in the previous time step and we only change the values for the current time step while iterating over the nodes.
\end{itemize}

\section{Pygrnd implementation of the Monte Carlo evaluation}\label{appendix 2}
In this section, we describe the Monte Carlo implementation of Pygrnd~\cite{pygrnd} for calculating the probabilities of the configurations. The function is {\tt monteCarloEvaluation} and it is in the module {\tt pygrnd.qc.probabilisticNetworks}.

\begin{itemize}
\item We start with the configuration $(0,\ldots, 0)$ for time step $t=0$.
\item For each new time step we iterate through all nodes.
\item If a node had state $1$ in the previous time step, we set it to $0$ for the next time step with probability $p^{\rm recover}_k$.
\item If a node had state $0$ in the previous time step, we set it to $1$ for the next time step with probability $p_k^{\rm fail}$.
\item Furthermore, we iterate over all ancestor nodes $m$ in the network that had state $1$ in the previous time step and we set the node $n$ to state $1$ in the next time step with probability $p_{m,n}^{\rm trigger}$.
\item After iterating over all nodes this gives us a configuration for time step $t+1$.
\item We repeat the procedure above until we obtain a configuration for the desired time step.
\item Once we reach the desired time step we store the configuration.
\item We repeat the procedure above to generate statistics on generated configurations.
\item We approximate the probability of a configuration by dividing the number of occurrences of a configuration by the total number of repetitions.
\end{itemize}

\end{appendices}

\end{document}